\documentclass[aps,pre,reprint,superscriptaddress]{revtex4-2}
\usepackage{amsmath,amssymb,amsthm,dcolumn}
\usepackage[utf8]{inputenc}
\usepackage{graphicx}
\usepackage{xcolor}
\usepackage{newtxtext}
\usepackage{placeins}
\usepackage[slantedGreek,cmintegrals]{newtxmath}
\usepackage{hyperref}
\hypersetup{colorlinks,linkcolor={blue!90!black},citecolor={blue!90!black},urlcolor={blue!90!black}}
\usepackage{stmaryrd}
\usepackage{enumitem}
\usepackage{accents}
\usepackage[compact]{titlesec}
\DeclareMathAlphabet{\mathcal}{OMS}{cmsy}{m}{n}
\definecolor{linkcolor}{HTML}{0000ff}
\hypersetup{colorlinks,linkcolor={linkcolor},citecolor={linkcolor},urlcolor={linkcolor}}
\UseRawInputEncoding

\makeatletter
\newcommand{\fmarki}{\ensuremath{\dagger}}
\newcommand{\fmarkii}{\ensuremath{\ast}}
\newcommand{\fmarkiii}{\ensuremath{\ddagger}}
    
\def\@fnsymbol#1{{\ifcase#1\or \fmarki\or \fmarkii\or \fmarkiii\else\@ctrerr\fi}}
\makeatother

\begin{document}
\title{A model for boundary-driven tissue morphogenesis}

\author{Daniel S. Alber}
\thanks{These authors contributed equally to this work.}
\affiliation{Department of Chemical and Biological Engineering, Princeton University, Princeton, NJ 08540}
\affiliation{The Lewis-Sigler Institute for Integrative Genomics, Princeton University, Princeton, NJ 08540}

\author{Shiheng Zhao}
\thanks{These authors contributed equally to this work.}
\affiliation{Max Planck Institute for the Physics of Complex Systems, N\"othnitzer Straße 38, 01187 Dresden, Germany}
\affiliation{Max Planck Institute of Molecular Cell Biology and Genetics, Pfotenhauerstraße 108, 01307 Dresden, Germany}
\affiliation{Center for Systems Biology Dresden, Pfotenhauerstraße 108, 01307 Dresden, Germany}

\author{Alexandre O. Jacinto}
\affiliation{Center for Computational Biology, Flatiron Institute, Simons Foundation, New York, NY 10010}

\author{Eric~F.~Wieschaus}
\affiliation{The Lewis-Sigler Institute for Integrative Genomics, Princeton University, Princeton, NJ 08540}
\affiliation{Department of Molecular Biology, Princeton University, Princeton, NJ 08540}

\author{Stanislav Y. Shvartsman}
\altaffiliation{To whom correspondence should be addressed: stas@princeton.edu, haas@pks.mpg.de.}
\affiliation{The Lewis-Sigler Institute for Integrative Genomics, Princeton University, Princeton, NJ 08540}
\affiliation{Center for Computational Biology, Flatiron Institute, Simons Foundation, New York, NY 10010}
\affiliation{Department of Molecular Biology, Princeton University, Princeton, NJ 08540}

\author{Pierre~A.~Haas}
\altaffiliation{To whom correspondence should be addressed: stas@princeton.edu, haas@pks.mpg.de.}
\affiliation{Max Planck Institute for the Physics of Complex Systems, N\"othnitzer Straße 38, 01187 Dresden, Germany}
\affiliation{Max Planck Institute of Molecular Cell Biology and Genetics, Pfotenhauerstraße 108, 01307 Dresden, Germany}
\affiliation{Center for Systems Biology Dresden, Pfotenhauerstraße 108, 01307 Dresden, Germany}

\renewcommand{\vec}[1]{\boldsymbol{#1}}
\renewcommand{\eqref}[1]{Eq.~(\ref{#1})}
\newcommand{\eqsref}[1]{Eqs.~(\ref{#1})}
\newcommand{\neqref}[1]{(\ref{#1})}
\newcommand{\Omegai}{\mathit{\Omega}}
\newcommand{\Gammai}{\mathit{\Gamma}}

\newcommand{\figref}[2]{(Fig.~\hyperref[#1]{\ref*{#1}#2})}
\newcommand{\figsrefl}[3]{(Figs.~\hyperref[#1]{\ref*{#1}#2--\ref*{#1}#3})}
\newcommand{\figsreft}[4]{(Figs.~\hyperref[#1]{\ref*{#1}#2} and \hyperref[#3]{\ref*{#3}#4})}
\newcommand{\figrefi}[2]{(Fig.~\hyperref[#1]{\ref*{#1}#2}, inset)}
\newcommand{\textfigref}[2]{Fig.~\hyperref[#1]{\ref*{#1}#2}}
\newcommand{\textfigrefi}[2]{Fig.~\hyperref[#1]{\ref*{#1}#2}, inset}
\newcommand{\sfigref}[2]{(Supplementary Fig.~\hyperref[#1]{\ref*{#1}#2})}
\newcommand{\mm}{\hyperref[mm]{\emph{Materials and Methods}}}

\date{\today}

\begin{abstract}
Tissue deformations during morphogenesis can be active, driven by internal processes, or passive, resulting from stresses applied at their boundaries. Here, we introduce the \emph{Drosophila} hindgut primordium as a model for studying boundary-driven tissue morphogenesis. We characterize its deformations and show that its complex shape changes can be a passive consequence of the deformations of the active regions of the embryo that surround it. First, we find an intermediate characteristic triangular shape in the 3D deformations of the hindgut. We construct a minimal model of the hindgut primordium as an elastic ring deformed by active midgut invagination and germ band extension on an ellipsoidal surface, which robustly captures the symmetry-breaking into this triangular shape. We then quantify the 3D kinematics of the tissue by a set of contours and discover that the hindgut deforms in two stages: an initial translation on the curved embryo surface followed by a rapid breaking of shape symmetry. We extend our model to show that the contour kinematics in both stages are consistent with our passive picture. Our results suggest that the role of in-plane deformations during hindgut morphogenesis is to translate the tissue to a region with anisotropic embryonic curvature and show that uniform boundary conditions are sufficient to generate the observed nonuniform shape change. Our work thus provides a possible explanation for the various characteristic shapes of blastopore-equivalents in different organisms and a framework for the mechanical emergence of global morphologies in complex developmental systems.
\end{abstract}

\maketitle

\section{\uppercase{Introduction}}
Morphogenesis can proceed through active mechanisms, which generate tissue deformations by changing cell behaviors within their bounds, or passive mechanisms, which generate deformations via external conditions imposed at their boundaries by neighboring tissues~\cite{green_resolving_2022}. The interplay between active and passive tissues is particularly important during gastrulation, when an embryo has multiple genetically patterned active tissues in addition to passive regions that all deform significantly and almost simultaneously~\cite{stern_gastrulation_2004}. 

Perhaps no developmental system is as well understood as the \emph{Drosophila melanogaster} embryo at the onset of gastrulation, composed of a monolayer of maternally patterned cells between an internal yolk and a vitelline membrane encapsulated by an ellipsoidal rigid chorion. At this stage, several canonical examples of active tissues that are genetically patterned to induce changes in cell shape or activity are undergoing morphogenesis: the posterior midgut (PMG), the ventral furrow (VF), and the germ band (GB)~(\citenum{stooke-vaughan_physical_2018,collinet_programmed_2021}, \textfigref{fig:Fig1}{B}). At the posterior pole, the PMG expresses the transcription factors Huckebein and Tailless~\cite{weigel_two_1990,harbecke_genes_1995,keenan_dynamics_2022} that signal through the GPCR ligand Fog to activate myosin and induce apical constriction and invagination of the posterior~\cite{costa_putative_1994, smits_design_2020, martin_physical_2020}. Similarly, a stripe of cells in the VF undergoes apical constriction and invaginates to form the mesoderm~\cite{leptin_cell_1990,sweeton_gastrulation_1991, oda_real-time_2001}. In addition to these out-of-plane deformations, the GB undergoes directed cell-cell rearrangements to converge and extend in-plane, pushing posterior tissue around the posterior pole onto the dorsal side of the embryo \cite{irvine_cell_1994,zallen_patterned_2004,bertet_myosin-dependent_2004, blankenship_multicellular_2006,kong_forces_2017}. 

While the deformations of these active tissues are striking, they are separated at their boundaries by a domain of cells that deforms no less dramatically, although it lacks obvious expression of genes regulating active deformation~\cite{keenan_dynamics_2022}. This circular domain will ultimately give rise to the hindgut and consists of approximately 450 cells expressing Brachyenteron (\emph{Drosophila} Brachyury). Brachyenteron defines a highly conserved signaling module specifying the posterior fates and gut formation in many organisms~\cite{technau_brachyury_2001, swalla_building_2006, lengyel_it_2002, wu_role_1998}. Homologs include T in mouse, No-tail in zebrafish, and XBra in \emph{Xenopus}, and are typically present at the lip of the blastopore-equivalent posterior internalization~(\citenum{bruce_brachyury_2020,anlas_studying_2021, schwaiger_ancestral_2021,hayata_expression_1999,gross_role_2001,yuan_differential_2020}, \textfigref{fig:Fig1}{A}). In \emph{Drosophila}, the domain is ring-shaped and located anterior to the PMG but posterior to the VF and GB~(\citenum{keenan_dynamics_2022}, \textfigref{fig:Fig1}{B}). Although Brachyenteron expression is ultimately required for cell-fate-specific differentiation of the hindgut, its elimination has no direct effect on the morphogenetic movements that occur at gastrulation~\cite{singer_drosophila_1996}. This raises the possibility that early morphogenesis in the hindgut is imposed by forces generated in the surrounding regions. 

\begin{figure*}[ht!]
\includegraphics[width=17.5 cm]{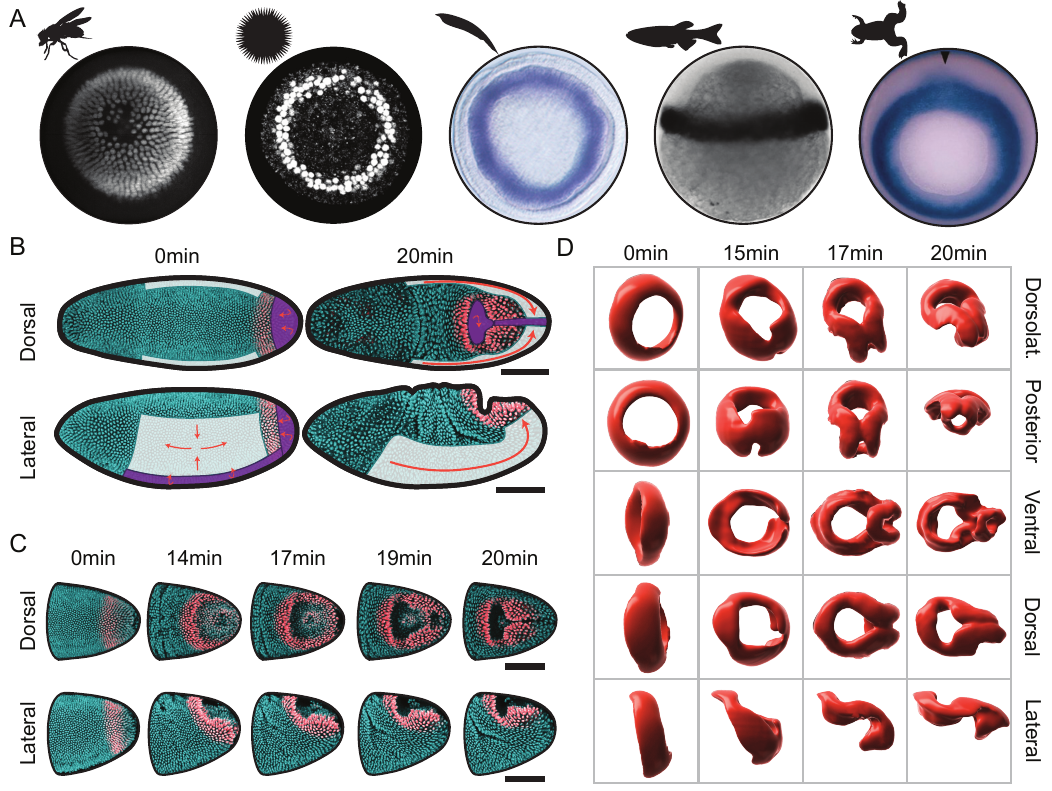}
\caption[Background]{{The hindgut primordium is bounded by active tissues and rapidly deforms in 15 minutes.} (A) \emph{brachyury} ortholog expression at the gastrula stage across the animal kingdom, in (left to right) the fruit fly \emph{Drosophila melanogaster}~\cite{keenan_dynamics_2022}, the sea urchin \emph{Lytechnius variegatus}~\cite{gross_role_2001}, the lancelet \emph{Branchiostoma floridae}~\cite{yuan_differential_2020}, the vertebrate ray-finned zebrafish \emph{Danio rerio}~\cite{bruce_brachyury_2020}, and the amphibian frog \emph{Xenopus laevis}~\cite{hayata_expression_1999}. (B)~Dorsal and lateral views of the blastoderm at the onset of gastrulation and 21 minutes later. The cyan signal is a nuclear reporter and the red signal is a nuclear reporter specific to the hindgut (\mm). The germ band, which undergoes in-plane convergent extension, is shaded in white. The ventral furrow and the posterior midgut undergo out-of-plane invagination and are shaded in purple. (C)~Dorsal (top) and lateral (bottom) views of  the deforming hindgut primordium at five timepoints, showing invagination of the posterior midgut as the hindgut deforms into its characteristic triangular shape. (D)~Different views of surface reconstructions of the hindgut primordium from fixed data at timepoints approximated by morphology. Scale bars: $100\,\text{\textmu m}$.}
\label{fig:Fig1}
\end{figure*}

Embryos provide numerous examples of active deformations in one region exerting forces on neighboring primordia, possibly contributing to their subsequent morphogenesis. Examples of such ``boundary-driven'' deformations include differential tissue growth driving brain gyrification~\cite{tallinen14} and vertebrate gut looping~\cite{savin_growth_2011, nerurkar_bmp_2017, gill_developmental_2024}, friction forces driving the first folding event of the zebrafish brain~\cite{smutny17,inman23} and myotome formation~\cite{tlili_shaping_2019}, and active contractility at the tissue boundary driving amniote embryogenesis~\cite{saadaoui_tensile_2020,caldarelli_self-organized_2024}. A large body of work has characterized the diverse cellular processes that arise in response to external forces~\cite{bailles_genetic_2019}, adjacent domains~\cite{lye_mechanical_2015,lye_polarised_2024}, and geometric constraints~\cite{fierling_embryo-scale_2022, lefebvre_geometric_2023, tang_collective_2022, gomez-galvez_scutoids_2018, gehrels_curvature_2023}, including at the level of individual contributions within a tissue exhibiting both active and passive cellular behaviors~\cite{brauns_geometric_2024,lye_polarised_2024, guo_evidence_2022}. However, explanations for global morphological changes of entire passive tissues in the necessary context of their active neighbors and geometric constraints have remained elusive. The \emph{Drosophila} hindgut primordium offers an ideal system to develop a framework for understanding the deformations of such a passive tissue.

In the following experiments, we derive a minimal physical model to investigate whether contributions from adjacent actively-deforming tissues and embryonic geometry are sufficient to explain the morphogenesis of the hindgut primordium. We couple our model with 3D imaging of live embryos to quantify the deformations of the hindgut primordium rigorously. We find that as the PMG, VF, and GB impose forces at the boundary of the hindgut primordium, the primordium itself deforms in a combination of in-plane and out-of-plane deformations, breaking the symmetry of its circular shape into a characteristic, intermediate triangular ``keyhole'' shape~\figref{fig:Fig1}{C,D}. By tracking cells, we reveal a two-stage process and show that the kinematics of both stages are consistent with the passive deformations expected from forces applied at its boundary by the extension of the germ band and the invagination of the midgut that surround it.

\section{\uppercase{Results}}
\subsection*{Description of hindgut deformation at discrete timepoints}
To visualize the deformations of the hindgut primordium, we used an endogenous fluorescent Brachyenteron protein reporter built on the LlamaTag system \cite{bothma_llamatags_2018, keenan_dynamics_2022} to identify the hindgut primordium combined with a standard fluorescently tagged histone nuclear reporter to visualize the entire embryo. Briefly, the LlamaTag system leverages maternally-deposited eGFP, which is imported to the nucleus upon the presence of a nanobody fused to the endogenous protein of interest (in our case, Brachyenteron). The deformations of the hindgut primordium were initially visualized using confocal microscopy (\mm). 

At the onset of gastrulation, the initially circular tissue deforms significantly in a few minutes with no divisions nor cell death, and limited, if any, cell rearrangements \figref{fig:Fig1}{C}. The ring initially rotates and translates along the surface of the embryo due to germ band extension (\textfigref{fig:Fig1}{C}, lateral views at 0--14 min) and partially internalizes due to contact with the apically constricting and invaginating posterior midgut (\textfigref{fig:Fig1}{C}, dorsal views at 0--14 min). After this initial phase, the ring rapidly deforms into a characteristic triangular ``keyhole'' shape (\textfigref{fig:Fig1}{C}, dorsal views at 17--20 min). 

To create a more detailed description of these intermediate shapes, wildtype embryos were fixed, optically cleared, and stained for Brachyenteron and cell membrane markers Armadillo and Discs large. To visualize the deforming hindgut in 3D at a high isotropic spatial resolution, embryos were imaged using light sheet microscopy (\mm) and staged based on their morphology. Surface reconstructions from the Brachyenteron immunofluorescence signal \figref{fig:Fig1}{D} reveal complex intermediate geometries in which the internalized ``keyhole'' and the triangular shape of the tissue remaining on the surface are more apparent.

\begin{figure*}[t!]
\centering
\includegraphics[width=17.5cm]{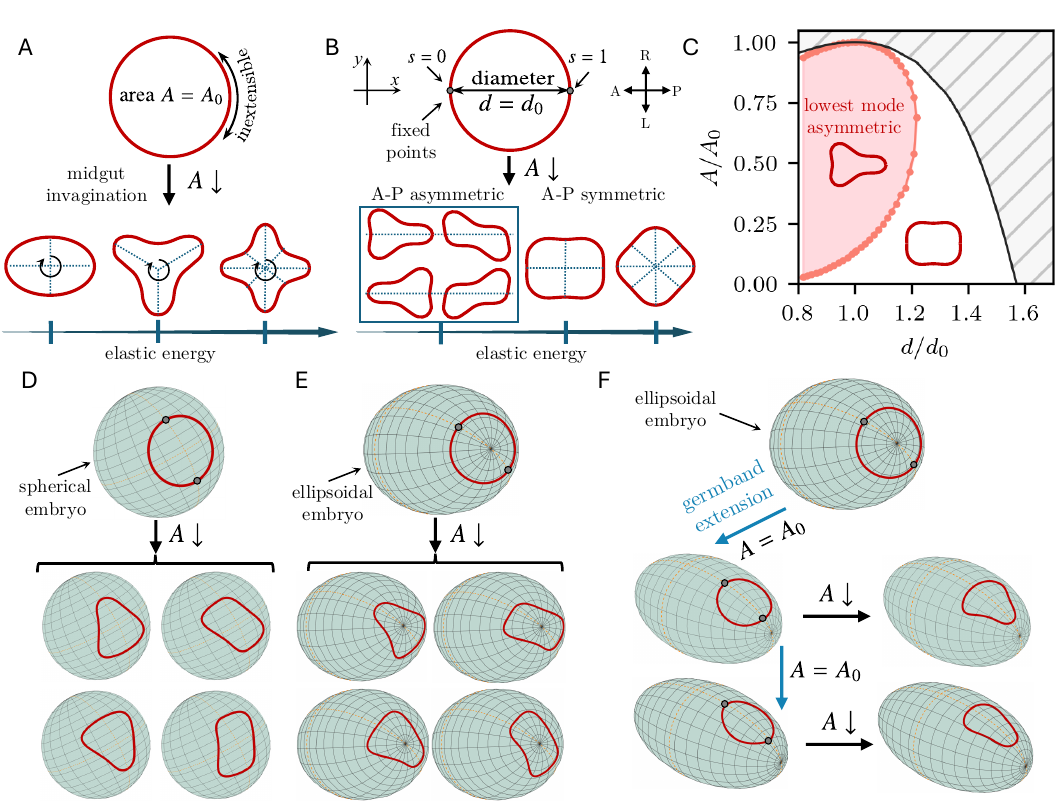}
\caption{{A minimal physical model reproduces the triangular keyhole shape of the primordial hindgut.} (A)~The primordial hindgut is modeled as a planar, inextensible elastic ring enclosing an initial area $A = A_0$. Invagination of the midgut reduces the enclosed area to $A$. The observed shape is the shape of lowest energy and symmetric~\cite{arreaga_elastica_2002,veerapaneni_analytical_2009}. Additional modes with higher energy also exist and have higher numbers of lobes~\cite{arreaga_elastica_2002,veerapaneni_analytical_2009}. (B)~The position of the germ band additionally sets the anteroposterior (AP) diameter $d$ of the ring, i.e., the distance between two diametrically opposite points at arclength positions $s=0$, $s=1$. For $d=d_0$, the four shapes of equal lowest energy are AP asymmetric, i.e., asymmetric about the $y$-axis , and include triangular shapes similar to the shape of the primordial hindgut. Additional symmetric and asymmetric shapes are possible as well, but are of higher energies (\mm). (C)~Phase diagram of the bifurcation from panel (B) in $(d, A)$ space: The AP asymmetric keyhole shape remains the lowest-energy mode in the shaded region of parameter space as $A$ (midgut invagination) and $d$ (germ band extension) vary. The hatched region is geometrically inaccessible to inextensible deformations. (D)~An inextensible elastic ring constrained to lie on a sphere breaks symmetry into one of four shapes with equal energies, analogous to the planar shapes in panel (B), as the area enclosed by the ring is reduced (midgut invagination) while a diameter is fixed (germ band extension). (E)~An elastic ring at the posterior pole of an ellipsoid embryo breaks symmetry similarly to the spherical case in panel (D). (F)~Symmetry-breaking of an elastic ring at the posterior pole of an ellipsoid after translation to the dorsal side (germ band extension) and reduction of the area enclosed by the ring (midgut invagination): Among the shapes in panels (D), (E), the gradient in curvature consistently selects the triangular shape with the orientation observed in the \emph{Drosophila} hindgut primordium.} 
\centering
\label{fig:Fig2}
\end{figure*}

\subsection*{Model of the symmetry-breaking of the hindgut}
We hypothesized that the shape changes of the hindgut primordium are the passive mechanical consequences of the deformations of the surrounding tissues. We therefore started by deriving a minimal theoretical model of hindgut morphogenesis. In this model, the hindgut primordium is skeletonized to a planar inextensible elastic ring $\mathscr{C}$ enclosing an area occupied by the posterior midgut. The ring is initially circular, of area $A=A_0$~\figref{fig:Fig2}{A}. As the midgut invaginates by apical constriction, the effective apical surface area of the tissue decreases, which reduces $A$ and deforms the ring. This deformation minimizes the bending energy of the ring,
\begin{align}
\mathcal{E}=\dfrac{1}{2}\oint_{\mathscr{C}}{\kappa(s)^2\,\mathrm{d}s},\label{eq:be}
\end{align}
where $s$ is arclength and $\kappa(s)$ is curvature, subject to the constraints imposing inextensibility and the area $A$ enclosed by $\mathscr{C}$ (\mm). This is a well-known mechanical problem~\cite{arreaga_elastica_2002,veerapaneni_analytical_2009}: The observed shape (of lowest energy) of an elastic ring enclosing a prescribed area is symmetric; higher modes of higher energy have higher numbers of lobes~\figref{fig:Fig2}{A}.

Our minimal model therefore needs one more constraint: The points at which the ring intersects the mid-sagittal cross section of the embryo cannot move freely, but their position is set at each timepoint by the progress of germ band extension. In the model, this fixes the distance $d$ between two diametrically opposite points on the ring, i.e., its anteroposterior (AP) diameter. The shape of the deformed ring minimizes its bending energy subject to these three constraints. The corresponding Euler--Lagrange equation is
\begin{subequations}
\begin{align}
\kappa''(s)+\dfrac{\kappa(s)^3}{2}-\lambda_0\kappa(s)+p=0,\label{eq:kappa}
\end{align}
where dashes denote differentiation with respect to $s$, and where $\lambda_0$ and $p$ are constants to be determined~(\mm). We complement this with the differential equations
\begin{align}
&\theta'(s)=\kappa(s),&& x'(s)=\cos{\theta(s)},&&y'(s)=\sin{\theta(s)},
\end{align}
\end{subequations}
for the tangent angle $\theta(s)$ with the AP axis, and the position $\bigl(x(s),y(s)\bigr)$ of a point on the ring~\figref{fig:Fig2}{B}. The boundary conditions fix the enclosed area to $A$ and the AP diameter to $d$ and impose the symmetry of the half-ring~\figref{fig:Fig2}{B}. They are (\mm)\begin{subequations}
\begin{align}
&\theta(0)=-\theta(1)=\dfrac{\pi}{2},\;x(0)=y(0)=y(1)=0,\;x(1)=d,
\end{align}
and
\begin{align}
\int_0^1{\left[y(s)\cos{\theta(s)}-x(s)\sin{\theta(s)}\right]\,\mathrm{d}s}=A.
\end{align}
\end{subequations}
We solve this boundary-value problem numerically (\mm) as $A$ is reduced for $d=d_0$, the initial diameter of the ring.  The lowest-energy shapes are now asymmetric about the $y$-axis, i.e., AP asymmetric~\figref{fig:Fig2}{B}. There are four shapes of equal energy, which include ``keyhole'' shapes reminiscent of the shape of the hindgut primordium. There are also AP symmetric shapes, but they have higher energy~\figref{fig:Fig2}{B}. More generally, $d$ and $A$ both vary as the germ band extends and the midgut invaginates. For inextensible deformations, part of $(d,A)$ space is geometrically excluded. The asymmetric shapes remain the lowest-energy shapes in a large part of the remaining $(d,A)$ space~\figref{fig:Fig2}{C}. This shows that the symmetry-breaking that can lead to triangular shapes is robust.

\subsection*{Selection of hindgut shape by embryonic curvature}
The four-fold degeneracy of the shapes of minimal energy in \textfigref{fig:Fig2}{B} raises the question: How does the embryo consistently select one of these orientations? To answer this, we extended our model of a planar ring to a spherical or ellipsoidal surface approximating the embryonic geometry. However, even for these simple curved surfaces, the equation analogous to \eqref{eq:kappa} becomes too complex to write down. Instead, we directly minimized the bending energy in \eqref{eq:be}, subject to the same constraints, for shapes approximated by a few Fourier terms (\mm). An elastic ring on a sphere~\figref{fig:Fig2}{D} or at the posterior pole of an ellipsoid~\figref{fig:Fig2}{E} still breaks symmetry as $A$ is reduced, but the shape degeneracy persists by symmetry. If, however, the ring translates off the posterior pole and onto one side of the ellipsoid (similarly to the translation of the hindgut primordium onto the dorsal side of the embryo due to germ band extension), then the curvature gradients eliminate the degeneracy and the ring selects a triangular shape in the same orientation as the shape of the hindgut primordium~\figref{fig:Fig2}{F}.

Our minimal model thus shows that uniform contraction, representing midgut invagination, is sufficient to explain the symmetry-breaking of the hindgut primordium, with the observed shape selected by the curvature of the embryonic surface. In particular, neither active deformations of the hindgut primordium, nor inhomogeneous forces from the extending germ band that surrounds it, nor heterogeneities in its passive mechanical properties are necessary to explain the triangular shape qualitatively.

\begin{figure}[t!]
\includegraphics[width=8.6 cm]{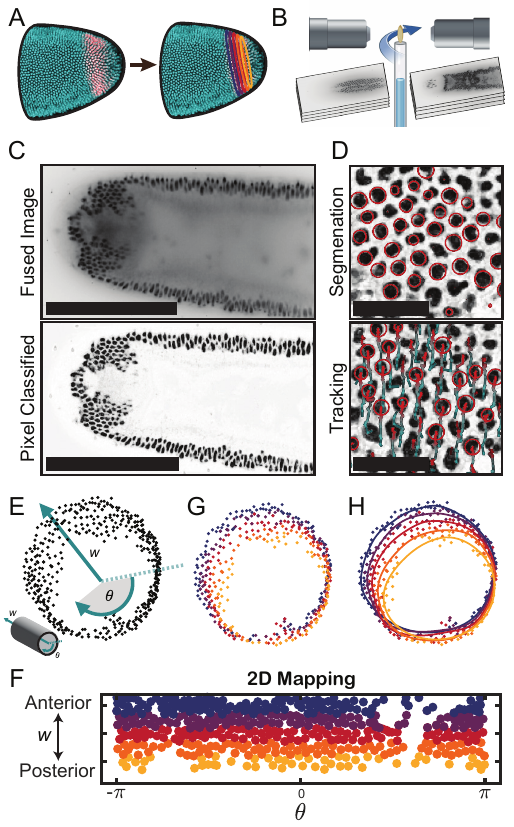}
\caption[Data analysis pipeline]{{Data analysis pipeline.} (A)~The analysis constructs a set of closed space curves (``contours'') that are initialized by positions of nuclei within the hindgut primordium and deform with it in time. (B)~Light sheet microscopy enables simultaneous imaging of both sides of embryos with fluorescent reporters for nuclei and hindgut. (C)~After image fusion and deconvolution (\mm), images are processed using a pixel classifier (ilastik, \citenum{berg_ilastik_2019}) to improve nuclear detection. Scale bars: $200\,\text{\textmu m}$. (D)~Nuclei within the hindgut primordium are segmented into spots (top); these spots are tracked semi-automatically using Mastodon~\cite{tinevez_mastodon_2022} (bottom) to generate a full track for each nucleus in the hindgut primordium. Scale bars: $20\,\text{\textmu m}$. (E)~Initial positions of nuclei at the blastoderm stage are mapped into cylindrical axial and angular coordinates $w,\theta$ (inset). (F)~Nuclei are binned into contours by their anteroposterior position $w$ in this 2D mapping. (G)~Initial nuclear positions from panel~(E) colored by the contour to which they are assigned from the binning in panel~(F). (H)~Contours are fitted using a sequence of splines that update at each timepoint as the nuclei move. Here, the initial contours are overlaid on from panel (G).}
\label{fig:FigDataAnalysis}
\end{figure}

\begin{figure*}[ht!]
\includegraphics[width=17.5 cm]{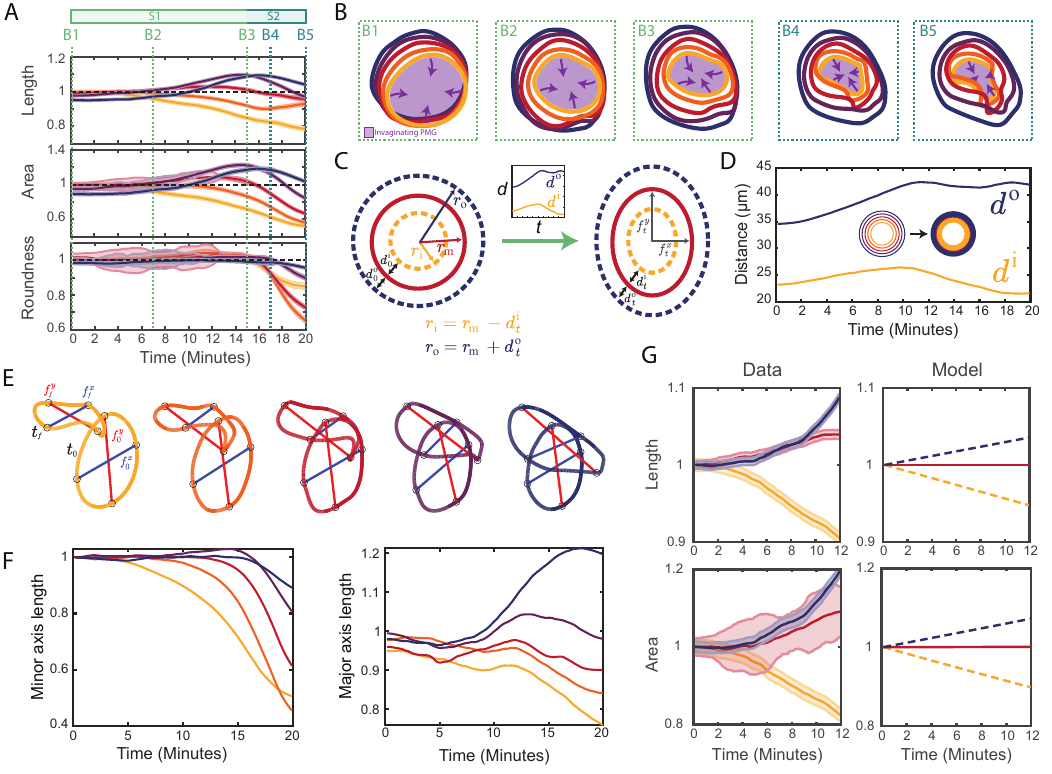}
\caption[The hindgut primordium deforms in two stages, shown for a representative embryo]{{The hindgut primordium deforms in two stages.} (A)~Shape metrics (contour length, enclosed area, and roundness) plotted against time for a representative embryo, colored by contour (innermost, yellow to outermost, blue). There are two stages: during stage S1 (green), all contours maintain their initial roundness and the lengths and areas of the inner and outer contours decrease and increase, respectively. From $t=15\,\text{min}$ onwards (stage S2, turqoise), the areas enclosed by all contours decrease and the roundness of all contours but the outermost one decreases sharply. Dashed lines, colored by stage, indicate timepoints B1--B5 used in panel (B). Error bars are determined from the standard deviation of a simulated error distribution (\mm). (B)~Contour shapes at the timepoints B1--B5 highlighted in panel (A). The violet shading indicates the invaginating posterior midgut (PMG). (C)~``Coupled-ring'' model of the deformation of circular contours into ellipses (\mm). At time $t$, the middle contour has semi-minor axis $\smash{f_t^x}$ and semi-major axis $\smash{f_t^y}$, and the initial distances $\smash{d_0^\text{i}}, \smash{d_0^\text{o}}$ from the middle to the inner and outer contours have changed to $d_t^\text{i}, d_t^\text{o}$, respectively (inset). (D)~Plot of the measured mean distances $\smash{d^\text{i}, d^\text{o}}$ from the middle to the innermost and outermost contours (\mm) against time. Inset: the contours define inner and outer rings used for calculating $\smash{d^\text{i}, d^\text{o}}$. (E)~Definition (\mm) of the minor (blue) and major (red) axis lengths $\smash{f_t^x, f_t^y}$, shown for each contour at the initial and final timepoints $\smash{t_0, t_f}$. (F)~Plots of the minor and major axis lengths or each contour, normalized by their initial lengths, against time. (G)~The ``coupled-ring'' model (right, \mm) sketched in panel (C) explains the kinematic behaviour of the inner and outer contours during stage S1 (left): If the length (top) or area (bottom) of the middle contour is constant (solid line), the model predicts (dashed lines) that the lengths or areas of the inner and outer contours decrease and increase, respectively, consistently with the data (left).}
\label{fig:Figure4}
\end{figure*}

\subsection*{Real-time kinematics inferred from live imaging}
To understand the kinematics of the hindgut primordium, we generated a series of closed space curves that we term ``contours''. Contours track the movement of nuclei within the hindgut primordium during the first 20 minutes of gastrulation \figref{fig:FigDataAnalysis}{A} and visualize the deformations of the hindgut primordium as threads on the surface of a fluid visualize its flow. First, we used light sheet microscopy to image (\mm) the deforming hindgut \figref{fig:FigDataAnalysis}{B}. We cooled the embryos to slow development, increasing the effective temporal resolution, and generated a 4D dataset with isotropic spatial resolution in one or two channels at a time resolution of 6--10s. After fusing and deconvolving images, we classified pixels using a standard tool~\cite{berg_ilastik_2019} to remove fluorescence from the yolk and beads used to register the images (\textfigref{fig:FigDataAnalysis}{C} and \mm). Pixel-classified images were segmented using a difference-of-Gaussians detector that approximates nuclei as 3D spheres (\citenum{tinevez_trackmate_2017,ershov_trackmate_2022}, \textfigref{fig:FigDataAnalysis}{D}). We tracked nuclei semi-automatically in Mastodon \cite{tinevez_mastodon_2022}, a tool built on the TrackMate \cite{tinevez_trackmate_2017,ershov_trackmate_2022} plugin for Fiji~\cite{schindelin_fiji_2012}. 
Each nuclear track was manually verified or corrected, resulting in approximately 500 tracks over approximately 100 timepoints~\figref{fig:FigDataAnalysis}{D}. We initialized contours by mapping the initial nuclear positions at the blastoderm stage into cylindrical coordinates~\figref{fig:FigDataAnalysis}{E}. Nuclei were binned into five groups based on their cylindrical axial coordinate $w$, corresponding to their embryonic anteroposterior positions~\figref{fig:FigDataAnalysis}{F}. Doing so divides the ring of the hindgut primordium into five slices~\figref{fig:FigDataAnalysis}{G}. Contours were fitted to each of these slices using a series of splines~\figref{fig:FigDataAnalysis}{H}. Contours were continually refitted using bins propagated from the initial assignments to capture the updated nuclear positions at subsequent timepoints (\mm), revealing the kinematics of the developing hindgut~(\hyperref[movieS1]{Movie S1}).

\subsection*{Two stages of hindgut morphogenesis}
To quantify the contour kinematics, we computed shape metrics at each timepoint and plot the normalized length, area, and roundness of each contour in \textfigref{fig:Figure4}{A} for a representative embryo. The length of the middle contour changes minimally over the first twenty minutes of gastrulation, which is consistent with the approximation of an inextensible midline and use of an elastic description (as opposed to an viscous description permitting cell rearrangements) in our physical model of the symmetry-breaking. Moreover, this quantification reveals that the deformation has two stages~\figref{fig:Figure4}{A}. 

During the first stage, the area and length of the outer and inner contours increase and decrease monotonically, respectively, while the area enclosed by the middle contour displays little to no change. The roundness of each contour remains close to unity, indicating uniform dilation and compression of the contours. Qualitatively, the shapes of all contours remain elliptical and begin to rotate and translate along the surface of the embryo as gastrulation begins~\figsrefl{fig:Figure4}{B1}{B3}. Towards the end of the first stage, apical constriction of the posterior midgut causes the areas enclosed by the contours to begin to decrease, starting with the innermost contour adjacent to the posterior midgut.

The second stage involves a sharp decrease of the roundness of all contours, with the outer contours remaining rounder than the middle and inner contours~\figref{fig:Figure4}{A}. As the contours move up and around the posterior pole~\figsreft{fig:Figure4}{B4}{fig:Figure4}{B5}, the midgut fully involutes and inverts, causing the areas enclosed by each contour to decrease~\figref{fig:Figure4}{A}. The contour lengths display more complex behavior, likely due to the out-of-plane deformations of the deforming hindgut. Interestingly, the inner contours, initially closer to the posterior, start to decrease in length, area, and roundness slightly before the outer contours. We computed the same metrics in terms of the position of the ring along the embryonic surface~\sfigref{sfig:SF_4}{A}, observing that all three shape metrics start to decrease when the ventralmost point of the contour passes the posterior pole~\sfigref{sfig:SF_4}{B}. This suggests that the delay results from different contours occupying similar regions of the embryo at slightly different times.

\subsection*{Minimal geometric model of the observed contour kinematics}
To explain the contrasting changes in the lengths and areas of the inner and outer contours during the first stage qualitatively, we introduced a minimal ``coupled--ring'' model describing an inner, middle, and outer contour (\textfigref{fig:Figure4}{C}, \mm). We hypothesized that the changes of the inner and outer contours are a consequence of the smaller deformations of the middle contour (which becomes slightly elliptical) and of the changes of its distance to the inner and outer contours. We therefore quantified (\mm) the mean distance between contours~\figref{fig:Figure4}{D} and their major (anteroposterior) and minor (left/right) axis lengths~\figsreft{fig:Figure4}{E}{fig:Figure4}{F}. To explain the relative behaviors of the inner and outer contours, we first modeled the length of the middle contour to be constant because of its lesser length change during stage S1. This predicts that the lengths of the inner and outer contours decrease and increase, respectively~\figref{fig:Figure4}{G}. Similarly, assuming that the area of the middle contour is constant, the model shows a decrease and increase of the inner and outer contour areas, respectively~\figref{fig:Figure4}{G}. The ``coupled-ring'' model thus captures the observed kinematics of the innermost and outermost contours. 

Interestingly, the major axis (i.e., the anteroposterior diameter) of the outermost contour has lengthened significantly by the end of the process~\figref{fig:Figure4}{F}, while the major axes of the other contours remain constant or shorten. At the same time, the shape of the outermost contour remains roundest~\figref{fig:Figure4}{A}. Only the middle and inner contours adopt the triangular shape that we have predicted in our minimal planar model. This is consistent with our model because shapes do not break symmetry if their anteroposterior diameter increases too much, as is the case for the blue outermost contour~\figref{fig:Fig2}{C}. 

\section{\uppercase{Discussion}}
Any developmental system comprised of both actively-deforming and passive tissues~\cite{tallinen14, savin_growth_2011, smutny17, inman23, saadaoui_tensile_2020, caldarelli_self-organized_2024} inevitably features deformations in boundary regions bridging actively-deforming neighbors. Such ``boundary-driven morphogenesis'' has proven difficult to understand, even at the level of kinematics, due to complex combinations of in-plane and out-of-plane deformations. This difficulty is compounded by the facts that boundary-driven and active morphogenesis can combine within the same tissue and that different combinations of passive and active cell behaviors can generate similar tissue deformations~\cite{brauns_geometric_2024, lye_mechanical_2015, lye_polarised_2024}. We have shown that our understanding of the morphogenesis of the \emph{Drosophila} hindgut primordium is consistent with a minimal model in which its complex deformations result solely from the forces exerted by its actively deforming neighboring tissues and the ellipsoidal geometry of the eggshell. Its dramatic change in shape, well-characterized neighboring tissues, and compatibility with well-established techniques for \emph{Drosophila} cell biology make the hindgut an ideal model for boundary-driven morphogenesis.

Previous work has described specific cellular processes in embryonic primordia ranging from the internalization of cells in the mesoderm~\cite{polyakov_passive_2014,guo_evidence_2022, mitrossilis_mechanotransductive_2017, chanet_actomyosin_2017} to biased cell rearrangements in the germ band~\cite{lye_polarised_2024,lefebvre_geometric_2023,brauns_geometric_2024}, proposing critical insights into how deformations may occur. Ultimately, fully understanding morphogenesis requires a more global approach that can integrate these individual findings. Here, we have taken such an approach that has allowed us to examine the full deformation of the hindgut primordium in its biological context. Our mechanism depends only on a uniform reduction of apical area by invagination of the posterior midgut and a uniform boundary condition from the germ band that translates the ring off the posterior pole. Movement of the ring to a region where the eggshell imposes anisotropic embryonic curvatures resolves the degeneracy of this symmetry-breaking and selects a triangular shape with proper orientation. Our minimal model absorbs these complex in-plane and out-of-plane deformations into simplified yet biologically relevant and measurable parameters, including the area enclosed by the tissue and its anteroposterior diameter. This paradigm will also be able to resolve which physical effects are likely to drive the observed global morphological changes in other developmental processes with complex boundary conditions. 

Although we have distilled the complex 3D shape of the hindgut that we observed in \textfigref{fig:Fig1}{D} into a triangular shape on the surface of the embryo, future work will need to understand the out-of-plane deformations of the internalized ``keyhole'' shape where the propagating ventral furrow meets the involuted midgut~\figref{fig:Fig1}{D}. In addition, we observed some in-plane stretching of the tissue between the contours in the anteroposterior direction, as evidenced by the changing intercontour distances~\figref{fig:Figure4}{D}. Further work will need to resolve the mechanical basis for this deformation within the hindgut. Continuum mechanical approaches~\cite{denberg_computing_2024} will enable elucidating the contributions of in-plane and out-of-plane boundary conditions from the neighboring active tissues to these and other characteristics of the hindgut shape. This will be aided by the rapid advances in techniques for measuring passive tissue properties~\cite{rauzi_embryo-scale_2015, cheikh_comprehensive_2023, goldner_evidence_2023, bevilacqua_high-resolution_2023, hart_myosin_2024, abbasi_cfm_2023}, perturbing cytoskeletal elements~\cite{guo_optogenetic_2022, herrera-perez_tissue_2023}, and machine-learning-assisted computer vision~\cite{moen_deep_2019,sugawara_tracking_2022,malin-mayor_automated_2023, nunley_nuclear_2024}, all of which will ultimately be used to populate a descriptive atlas of morphogenesis~\cite{mitchell_morphodynamic_nodate}. This approaching wave of data will couple to our framework to resolve mechanisms for global morphologies in development. 

More generally, by demonstrating the possible role of embryonic curvature in selecting the orientation of the triangular shape of the hindgut primordium, our work also offers an explanation for the effect of embryonic geometric constraints on the morphogenesis of other tissues. In many organisms, Brachyury is expressed at the lip of the blastopore or a similar invaginating structure~\cite{technau_brachyury_2001,anlas_studying_2021,wu_role_1998, meinhardt_radialsymmetric_2002} that deforms into various shapes depending on the organism. In some of these organisms, the blastopore lip appears as a constricting ring on a spherical embryo that fluidizes through cell rearrangements or oriented divisions, which can relieve stresses imposed at the boundaries through internal viscous dissipation~\cite{behrndt_forces_2012,campinho_tension-oriented_2013,martik_new_2017,mcclay_gastrulation_2020}. In some insects with more elongated embryos than those of \emph{Drosophila}, such as the medfly, germ band extension and posterior invagination differ, yet the lip of the posterior invagination also looks triangular as it moves off the posterior pole~\cite{strobl_two-level_2024}. In the beetle \emph{Tribolium castaneum}, the serosa undergoes epiboly through a mechanism separate from germ band extension and forms an intermediate triangular window on the ventral side of the ellipsoidal embryo~\cite{jain_regionalized_2020, munster_attachment_2019}. Using our framework to understand the mechanisms that drive the emergence of blastopore shapes will provide further insights into the evolution of the blastopore-to-primitive streak transition~\cite{stower_evolution_2017, chuai_reconstruction_2023}.

More physically, our triangular shape bifurcation expands the large body of work on constrained elastic lines in the plane and on curved surfaces~\cite{arreaga_elastica_2002,veerapaneni_analytical_2009,guven12,Giomi2012PRSA,Giomi2013SM,Hure2013,Box2020PRL,Kodio2020PRE,prasath21,shi2023} and related problems~\cite{audoly07,brun2015,Zakharov2018,Zakharov2020}. In this context, the shape-selection mechanism that we propose stresses the importance of anisotropic curvature for such bifurcations. The hindgut primordium and the ellipsoidal \emph{Drosophila} embryo more generally therefore provide a paradigm for mechanical bifurcations \emph{within} curved surfaces. Indeed, very recent work has shown that even the minimal instability that is Euler buckling changes fundamentally within general curved surfaces~\cite{zhao25}, but, compared to the well-understood instabilities \emph{of} curved surfaces~\cite{li2011PRL,tallinen14,budday15,stoop2015,miller2018prl,Tan2020review,Xu2022,tobasco2022,Wang2023}, these instabilities within curved surfaces remain mysterious.

\begin{acknowledgments}
We thank the members of the S.Y.S. and P.A.H. groups for helpful discussions and feedback. Imaging was performed with support from the Confocal Imaging Facility, a Nikon Center of Excellence, in the Department of Molecular Biology at Princeton University with assistance from Gary Laevsky. This work was supported by the NSF Graduate Research Fellowship under Grant \#DGE-2039656 (D.S.A.), NIGMS of the National Institutes of Health under award number R01GM134204 (S.Y.S.), and by the Max Planck Society~(S.Z. and P.A.H.).
\end{acknowledgments}

\appendix
\renewcommand{\thesection}{\textbf{APPENDIX \Alph{section}}}
\renewcommand{\theequation}{\Alph{section}\arabic{equation}}
\section{\uppercase{Materials \& Methods}}\label{mm}

\subsection{Imaging and tracking}
In order to visualize nuclei and identify hindgut progenitors, we generated a line containing the histone tag Histone H2B-RFP with the maternal ubiquitously-expressed GFP under the \textit{bicoid} promoter. Females from this stock were crossed with males containing the previously-generated Brachyenteron \mbox{LlamaTag~\cite{bothma_llamatags_2018,keenan_dynamics_2022}}. To generate the movie stills showing lateral and dorsal views of the deforming hindgut, embryos were manually dechorionated on double-sided tape before being immersed in halocarbon oil on custom filter slides and imaged using a Leica SP5 scanning confocal microscope. For tracking, embryos were manually dechorionated on double-sided tape and mounted in capillary tubes containing a solution of 1\% agarose with 1:200 diluted TetraSpeck $0.2\,\text{\textmu m}$ microspheres (ThermoFisher \#T7280). Imaging was performed on a Bruker/Luxendo MuVi-SPIM light-sheet microscope at $33.3\times$ magnification using two cameras mounted opposite each other and a rotating stage \figref{fig:FigDataAnalysis}{B}. Syncytial embryos were cooled to $18\,\text{\textdegree C}$ and full stacks were taken in the sagittal and frontal planes and in two channels (nuclei and Byn reporter) every 60 seconds to monitor the progression of development and designate cell identities. At the onset of gastrulation, defined as the onset of ventral furrow formation, imaging was switched to a single image stack acquired through the frontal plane (through the dorsoventral axis) in the histone reporter channel every 7.75 seconds at slice thicknesses of $1\,\text{\textmu m}$ to maximize temporal resolution. After 12--18 minutes, the imaging mode was switched back to the initial 2-channel, 2-angle mode to monitor further development. Embryos with visibly aberrant development or arrest were discarded from the dataset. Nuclei were tracked using Mastodon~(\citenum{tinevez_mastodon_2022} and \mm).
%\hyperref[details]{Appendix B}).

\subsection{Construction of contours from data} Raw tracks were smoothed by using an exponential moving average filter on each spatial dimension with a window size of 10 timepoints, or 80--110 seconds. Only nuclei that could be tracked through each timepoint were used. Approximately 5--10\% of nuclei, typically contained within the ventral midline in the ventral furrow, could not be  tracked reliably throughout the full movie. To initialize contours, positions of nuclei at the first timepoint were mapped into cylindrical coordinates \figref{fig:FigDataAnalysis}{E}. Positions were first normalized and then projected into eigenspace using a correlation matrix. Coordinates in eigenspace were converted to cylindrical coordinates, of which only the polar angle and the axial coordinate were used for mapping. Nuclei were binned into 5 bins based on their axial coordinates, corresponding to bands 2--3 nuclei wide to be used to fit contours. Each bin defined an initial contour identity, and these were propagated forward in time as nuclear positions changed.

To generate a contour at a given timepoint, points within the corresponding bin were sorted based on their initial azimuthal angle and their updated spatial coordinates were repeated three times to reduce edge effects. A cubic smoothing spline was applied to each dimension using the \texttt{csaps} function in \textsc{Matlab} (The MathWorks, Inc.) with a smoothing parameter of $0.01$. To extract a single closed contour, we iterated simultaneously in the forward and backward directions from the midpoint of the repeated array that contains the knots of these splines until these knots fell within a fixed tolerance of each other (which indicates completion of a full loop). A closed space curve was then obtained by joining the two knots and discarding knots outside of the interval containing the midpoint. All contours were resampled to generate space curves of 500 knots with constant arclength spacing.

\subsection{Calculation of shape metrics} The length $L$ of a contour was computed as the sum of the arc lengths of each spline within that contour. Its area $A$ was calculated by identifying its dorsalmost and ventralmost points to creating a line of bilateral symmetry. From this, the area was obtained as the (Riemann) sum of the lengths of line segments between corresponding points on either side of this midline multiplied by the distance between them. The roundness was defined to be $R = A/L^2$. In \textfigref{fig:Figure4}{}, each of these metrics was normalized by its value when the ventralmost point of the respective contour was located at a reference position. For this purpose, we first approximated the surface of the embryo by an ellipsoid with aspect ratio $2.5:1:1$, based on the aspect ratio $185\,\text{\textmu m}:92.5\,\text{\textmu m}:92.5\,\text{\textmu m}$ of the representative embryo used for \textfigref{fig:Figure4}{} measured using Fiji~\cite{schindelin_fiji_2012}. This reference position was then chosen to be the initial position of ventralmost point of the innermost (yellow) contour. To obtain metrics in terms of the positions of the contours~\figref{sfig:SF_4}{}, we reparameterized contours similarly by the positions of their ventralmost points along the arclength of a sagittal cross section of this ellipsoid, $s(\theta)=\mathrm{E}(\theta,\varepsilon)$, where $\mathrm{E}(\theta,\varepsilon)$ is the incomplete elliptic integral of the second kind, $\varepsilon=0.92$ is the eccentricity of this elliptical cross section, and $\theta$ is the polar angle measured from the anteroposterior axis.

\subsection{Physical models} Details of the derivation of the physical models are given in \hyperref[details]{Appendix B}.
\subsection{Additional experimental and image analysis methods} Further details of the experimental and image analysis methods are given in \hyperref[Suppmats]{Appendix C}.

\renewcommand{\thefigure}{S\arabic{figure}}
\setcounter{figure}{0}
\section{\uppercase{Physical Models}}\label{details}
\subsection{Derivation of the equation governing an inextensible elastic ring in the plane} 
As discussed in the main text, we begin by modeling the primordial hindgut as an inextensible elastic ring of length $\ell=2$ in the plane~\figref{sfig:SF_1}{A}, which we endow with Cartesian coordinates $(x,y)$. We parameterize the ring by its arclength $s$, so that a point on it has position $\vec{r}(s)=\bigl(x(s),y(s)\bigr)$. It is useful to consider half of the ring in the subsequent calculations and thus restrict to $s\in[0,1]$~\figref{sfig:SF_1}{A}. The deformed shape of the ring minimizes its bending energy subject to two constraints: (1) local inextensibility, $\|\vec{r}'(s)\|=1$, where the dash denotes differentiation with respect to $s$, and (2) the constraint that the half-ring enclose an area $A/2$, associated with midgut invagination, as discussed in the main text. The Lagrangian of the problem is
\begin{widetext}

\vspace{-12pt}
\begin{align}
\mathcal{L}=\dfrac{1}{2}\int_0^1{\|\vec{r}''(s)\|^2\,\mathrm{d}s}+\int_0^1{\lambda(s)\left[\|\vec{r}'(s)\|^2-1\right]\,\mathrm{d}s}+p\left(\int_0^1\frac{\vec{r}(s)\cdot\vec{n}(s)}{2}\,\mathrm{d}s - \dfrac{A}{2}\right),
\label{eq:L}
\end{align}
in which the first term is the bending energy, with $\|\vec{r}''(s)\|^2=\kappa(s)^2$, the squared curvature of the elastic line. In the other terms, $\lambda(s)$ is the Lagrange multiplier function associated with the inextensibility constraint, $p$ is the Lagrange multiplier that enforces area conservation, and $\vec{n}(s)$ is the unit normal to the ring. The tangent $\vec{t}(s)=\vec{r}'(s)$ and $\vec{n}(s)$ obey $\vec{t}'(s) = \kappa(s)\vec{n}(s)$ and $\vec{n}'(s)=-\kappa(s)\vec{t}(s)$. 

To be able to calculate the variation of the third term in \eqref{eq:L}, we need to note that $\vec{n}(s)\cdot\vec{n}(s) = 1 \Longrightarrow \delta \vec{n}(s)\cdot\vec{n}(s)=0$ and $\vec{n}(s)\cdot\vec{t}(s) = 0\Longrightarrow\delta\vec{n}(s)\cdot\vec{t}(s)=-\delta\vec{t}(s)\cdot\vec{n}(s)$, which imply $\delta\vec{n}(s) = -\left[\delta\vec{t}(s)\cdot\vec{n}(s)\right]\vec{t}(s)$. With this, and on integrating by parts several times, we obtain
\begin{align}
    \delta\mathcal{L} &= \int_0^1\left\{\left[\vec{t}'''(s)+2\lambda(s)\vec{t}'(s)+2\lambda'(s)\vec{t}(s) + \frac{p}{2}\vec{n}(s) + \frac{p}{2}\frac{\mathrm{d}}{\mathrm{d}s}\biggl(\left[\vec{r}(s)\cdot\vec{t}(s)\right]\vec{n}(s)\biggr)\right]\cdot\delta\vec{r}(s)+\delta\lambda(s)\left[\vec{r}'(s)\cdot\vec{r}'(s) - 1\right]\right\}\mathrm{d}s \nonumber\\
    &\qquad+ \left\llbracket\vec{r}''(s)\cdot\delta\vec{r}'(s) + \left(2\lambda(s)\vec{t}(s)-\vec{t}''(s)-\frac{p}{2}[\vec{r}(s)\cdot\vec{t}(s)]\vec{n}(s) \right)\cdot\delta\vec{r}(s)\right\rrbracket_0^1.
\label{eq:varL}
\end{align}
Hence

\vspace{-18pt}
\begin{align}
 \vec{t}'''(s)+2\lambda(s)\vec{t}'(s)+2\lambda'(s)\vec{t}(s) + p\vec{n}(s) + \frac{p}{2}\kappa(s)[\vec{r}(s)\cdot\vec{n}(s)]\vec{n}(s) - \frac{p}{2}\kappa(s)\left[\vec{r}(s)\cdot\vec{t}(s)\right]\vec{t}(s)=\vec{0}.\label{eq:EL}
\end{align}
Differentiating $\vec{t}'(s) = \kappa(s)\vec{n}(s)$ gives $\vec{t}''(s) = \kappa'(s)\vec{n}(s) - \kappa(s)^2\vec{t}(s)$ and $\vec{t}'''(s) = \kappa''(s)\vec{n}(s) - 3\kappa(s)\kappa'(s)\vec{t}(s) - \kappa(s)^3\vec{n}(s)$. With this, the normal and tangential components of \eqref{eq:EL} yield, respectively,
\begin{align}
&\kappa''(s)-\kappa(s)^3-2\lambda(s)\kappa(s)+p+\dfrac{p}{2}\kappa(s)\left[\vec{r}(s)\cdot\vec{n}(s)\right]=0,&& -3\kappa(s)\kappa'(s)+2\lambda'(s)-\dfrac{p}{2}\kappa(s)\left[\vec{r}(s)\cdot\vec{t}(s)\right]=0,\label{eq:ELC}
\end{align}
whence we obtain expressions for $\vec{r}(s)\cdot\vec{n}(s),\vec{r}(s)\cdot\vec{t}(s)$, which, in turn, give
\begin{align}
\vec{r}(s)=\dfrac{2}{p\kappa(s)}\left\{\left[\kappa''(s)-\kappa(s)^3+2\lambda(s)\kappa(s)+p\right]\vec{n}(s)+\left[-3\kappa(s)\kappa'(s)+2\lambda'(s)\right]\vec{t}(s)\right\}.
\end{align}
We can now differentiate this relation, use $\vec{r}'(s)=\vec{t}(s)$, and collect normal and tangential components to find
\begin{align}
    \kappa'''(s)\kappa(s)-\kappa'(s)\kappa''(s)+\kappa'(s)\kappa(s)^3-p\kappa'(s) &= 0,&\kappa''(s)+\dfrac{\kappa(s)^3}{2}-\lambda(s)\kappa(s)-\dfrac{p}{4}-\dfrac{\mathrm{d}}{\mathrm{d}s}\left(\dfrac{\lambda'(s)}{\kappa(s)}\right)&=0.
\label{eq:balance_nt}
\end{align}
The second equation gives the Lagrange multiplier function $\lambda(s)$ once $\kappa(s)$ is determined from the first, which rearranges to
\begin{align}
\dfrac{\mathrm{d}}{\mathrm{d}s}\left(\frac{\kappa''(s)}{\kappa(s)}\right)+\kappa(s)\kappa'(s)+p\dfrac{\mathrm{d}}{\mathrm{d}s}\left(\frac{1}{\kappa(s)}\right)=0.
\end{align}
On integrating, we finally obtain

\vspace{-18pt}
\begin{align}
\kappa''(s)+\dfrac{\kappa(s)^3}{2}-\lambda_0\kappa(s)+p=0,\label{eq:balance_n2}
\end{align}

\noindent where $\lambda_0$ is a constant of integration. [In deriving this equation, we have assumed that $\kappa(s)\neq0$, but, if $\kappa(s)=0$, then the final equation still holds by the first of \eqsref{eq:ELC}.] We are not aware of a previous derivation of this governing equation by direct variation of the Lagrangian, but the same equation has been obtained using the method of ``normal variation''~\cite{Arreaga2002,Diamant2011} or from the Kirchhoff rod equations~\cite{Kodio2020,Foster2022}. 

To write down the boundary conditions, we introduce the tangent angle $\theta(s)$~\figref{sfig:SF_1}{A}. Since $\theta'(s)=\kappa(s)$, \eqref{eq:balance_n2} is a third-order equation for $\theta(s)$. By definition, $x'(s) = \cos{\theta(s)}, y'(s) = \sin{\theta(s)}$, which are two first-order equations. Moreover, we are to determine two unknown constants, viz., $\lambda_0,p$. We therefore need $3+1+1+2=7$ boundary conditions. We impose~\figref{sfig:SF_1}{A}
\begin{align}
&\theta(0)=\dfrac{\pi}{2},&&\theta(1)=-\dfrac{\pi}{2},&&x(0)=0,&&x(1)=d,&&y(0)=y(1)=0,&&\int_0^1{\left[y(s)\cos{\theta(s)}-x(s)\sin{\theta(s)}\right]\mathrm{d}s}=A,
\end{align}
in which $d$ is the anterior-posterior extension of the ring (set by germband extension, as discussed in the main text), and the first two conditions ensure that the full ring does not have kinks at $s=0$ or $s=1$~\figref{sfig:SF_1}{A}. With these boundary conditions, the boundary terms in \eqref{eq:varL} vanish as required, because they imply $\delta\vec{r}(0)=\delta\vec{r}(1)=\delta\vec{r}'(0)=\delta\vec{r}'(1)=\vec{0}$.

\begin{figure*}[t]
\centering
\includegraphics[width=16cm]{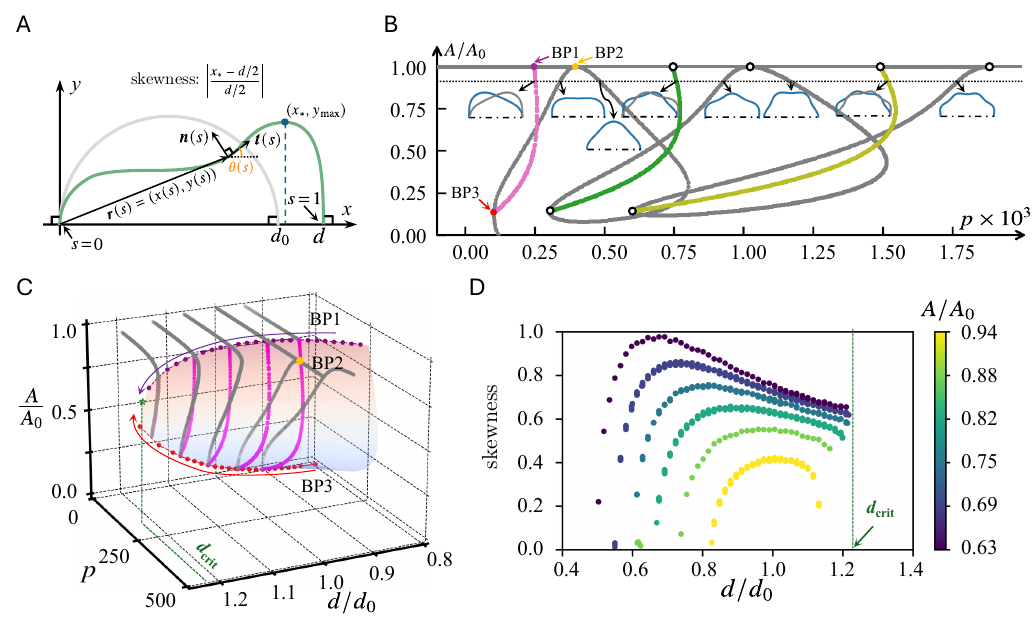}
\caption[Theory]{{Elastic ring in the plane: Bifurcation diagrams.} (A)~Deformation of an inextensible elastic half-ring of length $\ell/2=1$ in Cartesian axes $(x,y)$. A circular half-ring (gray line) encloses an area $A_0\equiv 1/\pi$ and has diameter $d_0\equiv 2/\pi$. The ring deforms as the area $A$ enclosed by the ring and its diameter $d$ are varied while minimizing its bending energy. The position of a point on the deformed half-ring (green line) is $\vec{r}(s)=(x(s),y(s))$, where $s\in[0,1]$ is arclength, so that $x(1)-x(0)=d,y(0)=y(1)$. The tangent and normal to the ring are $\vec{t}(s)$ and $\vec{n}(s)$, and the tangent angle is $\theta(s)$. Completing the shape of the half-ring into a full ring requires $\theta(0)=\pi/2,\theta(1)=-\pi/2$ (right angles emphasized). Inset equation: definition of the skewness of the deformed shape. (B)~Bifurcation diagram of the elastic half-ring for $d=d_0$ in $(p,A)$ space, where $p$ is pressure. Symmetric branches (gray lines) and asymmetric branches (colored lines) bifurcate off the undeformed branch $A=A_0$ at increasing values of $p$. Each asymmetric branch joins a symmetric branch. Circular markers are branch points; the first three branch points BP1, BP2, BP3 are highlighted. Insets: deformed shapes for $A/A_0=0.94$ on different branches; on asymmetric branches, two shapes reflected about $s=1/2$ are shown to emphasize the asymmetry. The first branch to bifurcate (magenta branch) is an asymmetric keyhole shape (similar to the shape of the posterior hindgut). (C)~Plot of the first asymmetric branch (shaded surface) in $(d,p,A)$ space. Magenta lines are sections at constant $d$; grey lines are the corresponding symmetric branches linked to these by the branch points BP1 and BP3. These points merge as $d\to d_\text{crit}\approx 1.22$; the first asymmetric branch ceases to exist for $d>d_\text{crit}$. The symmetric branches break up and BP2 disappears for $d\neq d_0$. (D) Plot of the skewness of the first asymmetric branch against $d$ for different $A$.}
\label{sfig:SF_1}
\end{figure*}

\subsection{Bifurcation diagrams} 
To study how a circular ring ($A=A_0\equiv 1/\pi,d=d_0\equiv 2/\pi$) deforms as $A$ and $d$ vary, we solve these equations using \texttt{AUTO-07p}~\cite{auto07p}. We first consider the case $d=d_0$, and plot the bifurcation diagram in $(p,A)$ space~\figref{sfig:SF_1}{B}. The first branch to bifurcate off the undeformed shape $A=A_0$, at branch point BP1, has asymmetric keyhole solutions akin to the shape of the hindgut (\textfigrefi{sfig:SF_1}{B}, as discussed in the main text). The first symmetric shapes only bifurcate at higher $p$, at branch point BP2. The asymmetric branch ends at another branch point, BP3, where it connects to this symmetric branch. As $p$ increases, additional symmetric and asymmetric branches bifurcate off $A=A_0$, with the asymmetric branches connecting to the symmetric branches that snake around them~\figref{sfig:SF_1}{B}.

As $d$ varies~\figref{sfig:SF_1}{C}, BP1 moves, in $(d,p,A)$ space, to $A<A_0$, but the asymmetric branch continues to be the lowest branch where it exists. Indeed, at $d=d_\text{crit}\approx 1.22$, BP1 merges with BP3, and this asymmetric branch ceases to exist. For $d\neq d_0$, branch point BP2 disappears, and the symmetric branches that merge there at $d=d_0$ break up~\figref{sfig:SF_1}{C}. As discussed in the main text, this bifurcation diagram shows that the asymmetric branch is the lowest (and hence the observed branch) for a range for values around $d=d_0$.

We quantify the asymmetry of these shapes by computing the skewness of the corresponding shapes~\figref{sfig:SF_1}{A}. The plot of skewness against $d$~\figref{sfig:SF_1}{D} emphasizes how, for each value of $A<A_0$, solutions of non-zero skewness exist in a range of values of $d<d_\text{crit}$. 

\subsection{Extended model: Mechanics of an inextensible elastic ring on a curved surface} As discussed in the main text, we extend our model to describe an inextensible elastic ring on a curved surface. Similarly to the plane case, the shape of a curve $\Gammai$ of prescribed length and enclosed area, confined to lie on a surface $\Omegai$, is determined by minimizing the bending energy of the curve,
\begin{equation}
    E = \frac{1}{2}\oint_\Gammai {\kappa(s)^2\,\mathrm{d}s},
\end{equation}
where $\kappa(s)$ is the total curvature of $\Gammai$ and $s$ is its arclength, subject to the constraints of prescribed length, prescribed surface area, and the constraint of the curve lying on $\Omegai$.

\begin{figure*}[t]
\centering
\includegraphics[width=11cm]{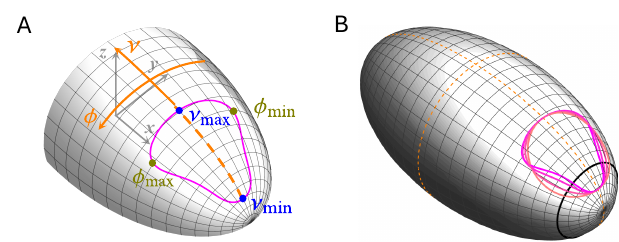}
\caption[Theory]{{Extended model: Elastic ring on a curved surface.} (A) Geometry of a closed curve $\Gammai_\Omegai$ on an ellipsoid with semi-axes $a,b=c$, in terms of the polar angles $\nu,\phi$. The position of the curve is defined by anterior-most and posterior-most points $\phi=0$, $\nu=\nu_{\min},\nu_{\max}$, and the curve encloses the area $\phi_{\min}(\nu)\leqslant\phi\leqslant\phi_{\max}(\nu)$ for $\nu_{\min}\leqslant\nu\leqslant\nu_{\max}$. (B) Example result of the extended model: An elastic circular ring moved from the posterior pole (initial black ring) to the dorsal side breaks symmetry when the area $A$ enclosed by the ring is reduced. Parameter values: $a=2,b=c=1$; initial circle: $\nu=0.5$.}
\label{sfig:SF_1b}
\end{figure*}

We now specialize to ellipsoidal surfaces~\figref{sfig:SF_1b}{A}, which include spherical surfaces as a special case, and begin by imposing the condition $\Gammai\subset\Omegai$ by choosing an explicit parametrization of $\Gamma$. In a Cartesian coordinate system with position vector $\vec{r}=(x,y,z)$, the position of a point on the surface of an ellipsoid can be written as
\begin{align}
    &x(\nu) = a \cos{\nu}, &&y(\nu,\phi) = b \cos{\nu}\sin{\phi}, &&z(\nu,\phi) = c \cos{\nu}\cos{\phi}, 
\end{align}
where $a,b,c$ are the semi-axes of the ellipsoid, and $\nu\in[0,\pi]$ and $\phi\in[0,2\pi]$ are its polar and azimuthal angles, respectively. To impose $\Gammai\subset\Omegai$, we choose a parametrization 
\begin{align} 
\Gammai_\Omegai(\tau)\colon \tau \to \bigl\{ x\bigl(\nu(\tau),\phi(\tau)\bigr),y\bigl(\nu(\tau),\phi(\tau)\bigr),z\bigl(\nu(\tau),\phi(\tau)\bigr)\bigr\},
\end{align}
where $\tau\in[0,T]$ parametrizes the curve, for some $T>0$. The total squared curvature of $\Gammai$ is now
\begin{align}
    \kappa (\tau)^2 = \frac{\left[z''(\tau)y'(\tau)-y''(\tau)z'(\tau)\right]^2+\left[x''(\tau)z'(\tau)-z''(\tau)x'(\tau)\right]^2+\left[x''(\tau)y'(\tau)-y''(\tau)x'(\tau)\right]^2}{\left[x'(\tau)^2+y'(\tau)^2+z'(\tau)^2\right]^{3/2}},
\end{align}
where dashes now denote differentiation with respect to $\tau$. The length of the curve and the area of the enclosed region are
\begin{align}
    &\oint_{\Gammai_\Omegai} \mathrm{d}s
      = \int_0^{T}{\left[x'(\tau)^2+y'(\tau)^2+z'(\tau)^2\right]^{1/2}\,\mathrm{d}\tau}, && \oiint_{\Gammai_\Omegai}{\mathrm{d}A}
    =\int_{\nu_{\min}}^{\nu_{\max}}
    \int_{\phi_{\min}(\nu)}^{\phi_{\max}(\nu)} \left\| \frac{\partial \mathbf{r}}{\partial\nu} \times \frac{\partial\mathbf{r}}{\partial\phi}  \right\| \mathrm{d}\phi \, \mathrm{d}\nu ,
\end{align}
respectively, where the values $\phi_{\min}(\nu),\phi_{\max}(\nu)$ and $\nu_{\min},\nu_{\max}$ are associated with the shape of the curve~\figref{sfig:SF_1b}{A}. A curve $\Gammai$ of prescribed length $L$ and prescribed enclosed area $A$ on $\Omegai$ now extremizes the functional
\begin{equation}
\label{energyfunctional}
\begin{aligned}
    \mathcal{F} &= \oint_{\Gammai_\Omegai}{\!\kappa(\tau)^2s'(\tau)\,\mathrm{d}\tau} + \lambda_1 \left ( \oiint_{\Gammai_\Omegai}{\mathrm{d}A}-A \right)+ \lambda_2 \left(\oint_{\Gammai_\Omegai}{\mathrm{d}s}-L\right),
\end{aligned}
\end{equation}
where $\smash{s'(\tau)=\left[x'(\tau)^2+y'(\tau)^2+z'(\tau)^2\right]^{1/2}}$. The first term on the right-hand side is the bending energy of the curve (up to a factor of $1/2$), and the second and third terms impose the area and length constraints by Lagrange multipliers $\lambda_1,\lambda_2$. 

Variation of $\mathcal{F}$ with respect to the functions $\nu(\tau),\phi(\tau)$ determines, in principle, a boundary-value problem that sets the shape of the energy-minimizing curve. However, it turns out that the calculations determining this boundary-value problem are prohibitively cumbersome. We therefore take a different numerical approach: For simplicity, we consider a curve that can be expressed in terms of a few Fourier coefficients, viz.,
\begin{align}
        &\nu(\tau) = b_0+b_1 \cos{\tau}, &&\phi(t) = a_1 \sin{\tau}+ a_2\sin{2\tau}+a_3\sin{3\tau}.
\end{align}
In what follows, we will determine $b_0,b_1$ by imposing the positions $\nu_{\min},\nu_{\max}$ of the anterior-most and posterior-most points along the curve on the ellipsoid~\figref{sfig:SF_1b}{A}. The length and area constraints then give two relations between the remaining three parameters $a_1,a_2,a_3$, and their values are finally set by minimizing the elastic bending energy subject to these relations numerically. This numerical minimization is performed using the \texttt{NMinimize} function of \textsc{Mathematica} (Wolfram, Inc.).

In \textfigref{sfig:SF_1b}{B}, we illustrate this approach, starting from a circle of length $A_0$ and circumference $L_0$. We move this curve along the ellipsoid to a different position on its dorsal side, and minimize $\mathcal{F}$ for $L=L_0$ and $A<A_0$. Reducing the imposed area enclosed by the curve in this way, we find a symmetry breaking leading to keyhole shapes of the hindgut, as discussed in more detail in the main text.

\begin{figure*}[t]
\centering
\includegraphics[width=15cm]{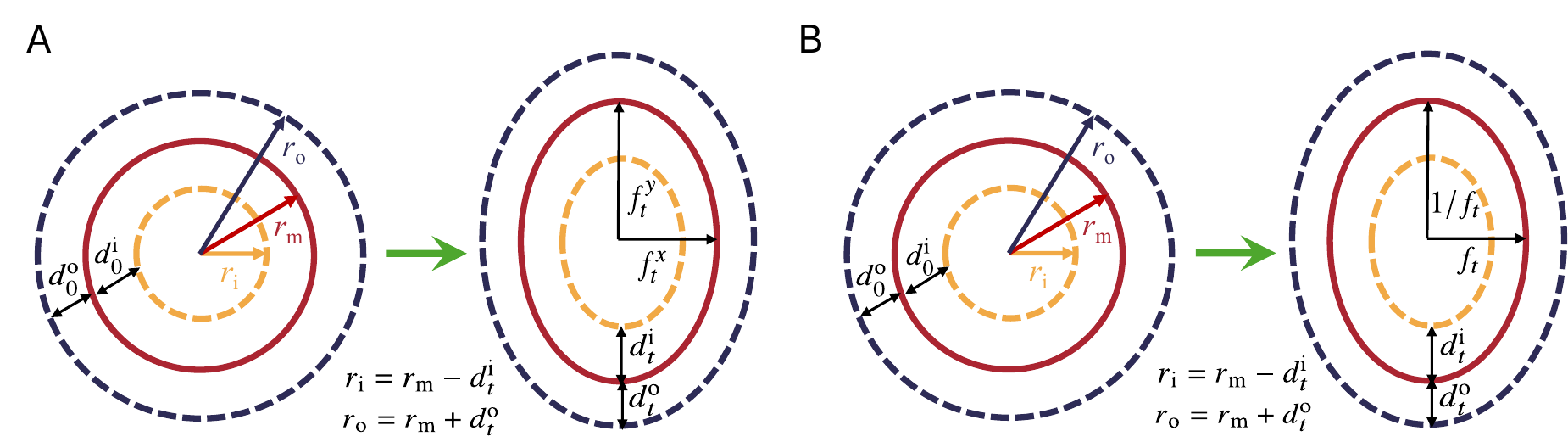}
\caption[Theory]{{``Coupled-ring'' model.} The model consists of three concentric circles in the plane that have initial radii $r_\text{i},r_\text{m},r_\text{o}$, which satisfy $\smash{r_\text{o} - r_\text{m} = r_\text{m} - r_\text{i} = d_0^\text{o}=d_0^\text{i}}$. As these rings deform into ellipses with semi-minor axis $f_t^x$ and semi-major axis $f_t^y$ at time $t$, the distances between the rings change to $\smash{d_t^\text{o}, d_t^\text{i}}$, respectively. We analyze two cases, in which (A) the length of the middle is conserved and (B) the area of the middle ring is conserved.}
\label{sfig:SF_1c}
\end{figure*}

\subsection{``Coupled-ring'' model} We construct a model of coupled rings that deform into ellipses to explain the observed length and area changes of the inner and outer contours, discussed in the main text, qualitatively. Here we provide the details of the derivation of the model. We consider three concentric circles with initial lengths and areas 
\begin{subequations}
\begin{align}
    &L_0^\text{i} = 2\pi\bigl(r_\text{m} - d_0^\text{i}\bigr), &&L_0^\text{m} = 2\pi r_\text{m}, && L_0^\text{o} = 2\pi\bigl(r_\text{m} + d_0^\text{o}\bigr),\\
    &A_0^\text{i}= \pi\bigl(r_\text{m} - d_0^\text{i}\bigr)^2, &&A_0^\text{m} =\pi r_\text{m}^2, && A_0^\text{o} = \pi\bigl(r_\text{m} + d_0^\text{o}\bigr)^2,
\end{align}
\end{subequations}
where sub- or superscripts i, m, o refer to the inner, middle, and outer rings, respectively and where $\smash{r_\text{m}}$ is the initial radius of the middle ring and $\smash{d_0^\text{i}, d_0^\text{o}}$ denote the initial distances of the middle ring to the inner and outer rings, respectively. We assume that, as the inner rings deforms into an ellipse with semi-minor and semi-major axes $\smash{f_t^x, f_t^y}$, respectively, the inner and outer rings deform into ellipses with axes $\smash{f_t^x-d_t^\text{i},f_t^y-d_t^\text{i}}$ and $\smash{f_t^x+d_t^\text{o},f_t^y+d_t^\text{o}}$, respectively, where $\smash{d_t^\text{i},d_t^\text{o}}$ are the distances between rings at time $t$. Hence the eccentricities of the ellipses are
\begin{align}
&\varepsilon_t^\text{i}=\left[1-\left(\dfrac{f_t^x-d_t^\text{i}}{f_t^y-d_t^\text{i}}\right)^2\right]^{1/2},&&\varepsilon_t^\text{m}=\left[1-\left(\dfrac{f_t^x}{f_t^y}\right)^2\right]^{1/2},&&\varepsilon_t^\text{o}=\left[1-\left(\dfrac{f_t^x+d_t^\text{o}}{f_t^y+d_t^\text{o}}\right)^2\right]^{1/2}.
\end{align}
The lengths and areas of the rings at time $t$ are therefore
\begin{subequations}
\begin{align}
    &L_t^\text{i} = 4\bigl(f_t^y - d_t^i\bigr)\mathrm{E}(\varepsilon_t^{\text{i}}), &&L_t^{\text{m}} = 4f_t^y\mathrm{E}(\varepsilon_t^{\text{m}}), && L_t^{\text{o}} = 4\bigl(f_t^y + d_t^o\bigr)\mathrm{E}(\varepsilon_t^{\text{o}}),\\
    &A_t^\text{i}= \pi\bigl(f_t^y - d_t^i\bigr)\bigl(f_t^x - d_t^i\bigr), && A_t^{\text{m}} =\pi f_t^xf_t^y, && A_t^{\text{o}} = \pi\bigl(f_t^y + d_t^o\bigr)\bigl(f_t^x + d_t^o\bigr),
\end{align}
\end{subequations}
in which $\mathrm{E}(\varepsilon)$ is the complete elliptic integral of the second kind. We now consider two cases: In the first case, the length of the middle ring is conserved~\figref{sfig:SF_1c}{A}. The condition $\smash{L_t^\text{m}=L_0^\text{m}}$ is an equation for $\smash{f_t^x}$ given $\smash{f_t^y}$ which can be solved numerically. In the second case, the area of the middle ring is conserved~\figref{sfig:SF_1c}{B}, and the condition $\smash{A_t^\text{m}=A_0^\text{m}}$ yields $\smash{f_t^x= r_\text{m}^2/f_t^y}$.

\begin{figure*}[ht!]
\centering
\includegraphics[width=16 cm]{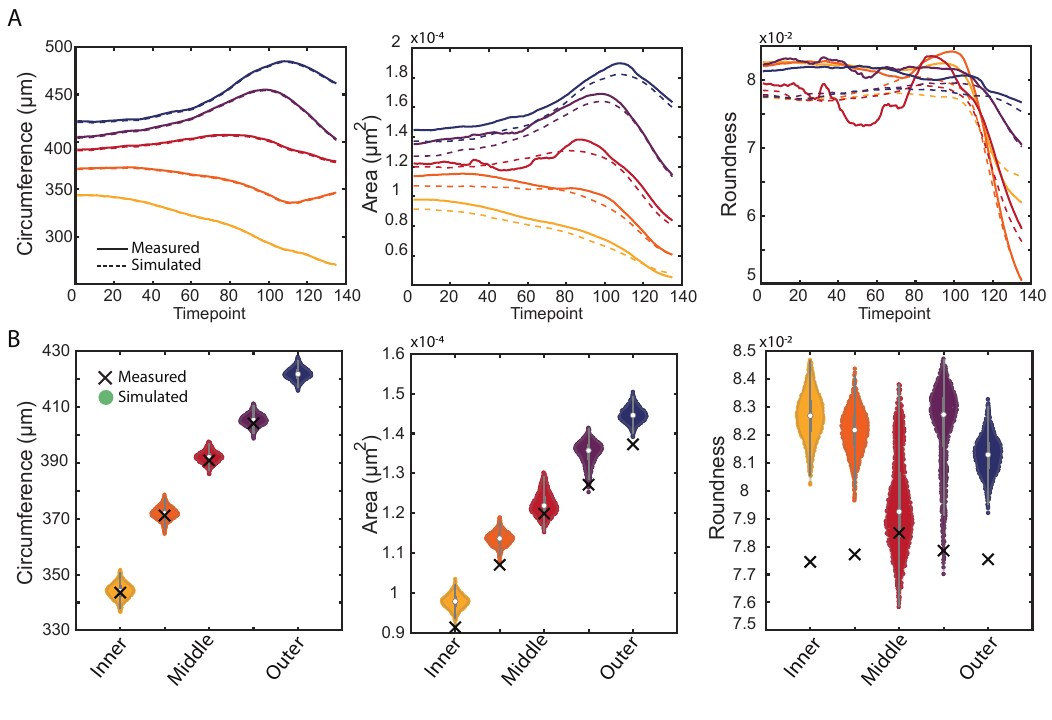}
\caption[Simulated error]{{Analysis of the simulated error.} (A) Plots of the means of the simulated contour circumferences, areas, and roundnesses (dashed lines) against time, colored by contour. They show the same qualitative behaviour as the measured values (solid lines). (B) Comparison of the distributions of the simulated circumferences, areas, and roundnesses to their measured values ($\times$), shown for each of the five contours and for the first timepoint.}
\label{sfig:SF_3}
\end{figure*}

\end{widetext}

We fit linear approximations to the experimental dynamics of $f_t^y,d_t^\text{i},d_t^\text{o}$~\figsreft{fig:Figure4}{D}{fig:Figure4}{F}. With these, $L_t^\text{i},L_t^\text{m},L_t^\text{o},A_t^\text{i},A_t^\text{m},A_t^\text{o}$ are determined in both cases, and we find that the length ratios $\smash{L_t^\text{o}/L_0^\text{o}}$ and $\smash{L_t^\text{i}/L_0^\text{i}}$ increase and decrease, respectively, in the first case, and that the area ratios $\smash{A_t^{\text{o}}/A_0^{\text{o}},A_t^{\text{i}}/A_0^{\text{i}}}$ increase and decrease, respectively, in the second case. As discussed in the main text, these results are consistent with the experimental observations~\figref{fig:Figure4}{G}.

\section{\uppercase{Additional experimental and image analysis methods}}\label{Suppmats}

\subsection{Image fusion and deconvolution} The imaging dataset for each embryo contains three sets of images, one for each of the three temporal imaging stages discussed in the \hyperref[mm]{Materials and Methods} section of the main text. The subset of this dataset used for contour generation includes the last timepoint in the first stage to identify the hindgut progenitors, the entire second stage involving rapid imaging of hindgut deformation, and the first timepoint in the last stage. Each stack contains 180--200 slices at a resolution of $2048\times 2048$ pixels and a spatial resolution of $0.195\,\text{\textmu m} \times 0.195\,\text{\textmu m} \times 1\,\text{\textmu m}$, totaling approximately $80\,000$ images for a given 130-timepoint dataset. These images were fused and deconvolved using the BigStitcher~\cite{horl_bigstitcher_2019} Fiji~\cite{schindelin_fiji_2012} plugin. Embedded fluorescent beads were used to register the images and generate point spread functions. The output was a series of image stacks comprising approximately $40\,000$ images containing the first and last timepoint in two channels and middle timepoints only in the histone channel. 

\subsection{Surface visualizations} To construct surface visualizations, wildtype (Oregon R) embryos were collected and fixed using heat fixations as detailed in Ref.~\cite{keenan_dynamics_2022}. Embryos were stained with antibodies against Brachyenteron and Discs-large as a cell surface marker. Embryos were imaged by light-sheet microscopy using the live imaging and image processing protocols described above and in the \hyperref[mm]{Materials and Methods} section of the main text. Surfaces were constructed using \textsc{Imaris} (Oxford Instruments, Inc.).

\subsection{Nuclear detection and tracking} In preparation for nuclear detection, images were downsampled by 2-4$\times$ in each dimension and pixel-classified using ilastik~\cite{berg_ilastik_2019}. The output of the pixel classifier is an image with identical resolution in which each pixel value corresponds to the probability of it belonging to the nuclear class. The pixel-classified images were imported into Mastodon~\cite{tinevez_mastodon_2022}, a Fiji~\cite{schindelin_fiji_2012} plugin built on the popular TrackMate~\cite{tinevez_trackmate_2017} framework. Nuclei were detected using a difference-of-gaussians detector. The initial two-channel timepoint containing information on the Brachyenteron reporter was used to identify 350--500 nuclei within the hindgut primordium. These nuclei were tracked semi-automatedly across approximately 130 timepoints using Mastodon with manual corrections and interventions. Each nuclear track was verified manually, culminating in approximately $50\,000$ annotations for one embryo. Particular attention was given to nuclei in internalizing regions of the tissue, where light scatter from the yolk resulted in diminished image quality and impaired automated detection and tracking. 

\subsection{Generation of simulated errors}\label{ssec:ContourError}
The semi-automated tracking means that the actual positions of the nuclear centroids could differ from the tracked points by up to a nuclear diameter. As a result, we determined the expected contribution of detection error by 1000 samplings from a uniform distribution of points within a nuclear diameter from the observed location for each nucleus at each timepoint. At each iteration, a simulated contour was generated using the approach described in the \hyperref[mm]{Materials and Methods} section of the main text, resulting in 1000 simulated contours for each of the 5 contours at each of the 135 timepoints. Lengths, areas, and roundness of these simulated contours were calculated. The means of each of these simulated metrics for the representative embryo analyyzed in \textfigref{fig:Figure4}{} of the main text are shown in \textfigref{sfig:SF_3}{A} and reproduce the measured values qualitatively. The distribution of the simulated metrics is compared, for one timepoint, to the measured values in \textfigref{sfig:SF_3}{B}. The errors for each metric were taken to be the standard deviation of the respective simulated metrics for each contour at each timepoint.

\subsection{Quantification of distances between contours} The distance between contours at each timepoint was taken to be the mean of the distances between each point on one contour and the closest point on the next contour. The mean distances between the middle contour and the outermost and innermost contours are plotted against time in \textfigref{fig:Figure4}{D}. At each timepoint, the minor axis length $f_t^x$ of each contour was defined to be the Euclidean distance between the two points located halfway between the initially dorsalmost and ventralmost points of the contour, as measured by their projection onto the anteroposterior axis \figref{fig:Figure4}{E}. The major axis length $f_t^y$ was taken to be the Euclidean distance between the initially dorsalmost and ventralmost points \figref{fig:Figure4}{E}. These lengths were computed for each contour, although only the innermost, middle, and outermost (yellow, red, and purple) contours were used for the ``coupled-ring'' model.

\onecolumngrid\newpage
\section{\uppercase{Supplemental materials}}
\subsection{Supplemental Figure}
\vspace{-15pt}
\begin{figure*}[h!]
\centering
\includegraphics[width=16cm]{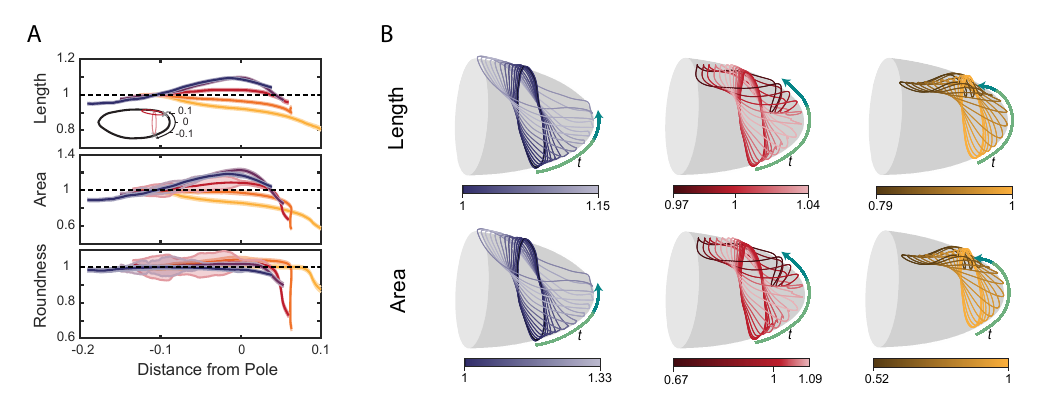}\vspace{-15pt}
\caption[Contour change based on position on embryo]{{Contour dynamics based on hindgut position.}  
(A) Metrics from \textfigref{fig:Figure4}{A} plotted against the distance, along the surface of the embryo, of the ventralmost point of each contour; the position is normalized by arclength (inset).  (B) Shapes of the outermost, middle, and innermost contours (blue, red, and yellow contours) plotted at 90 second intervals on top of a gray ellipsoid representing the surface of the embryo. The color shade of each contour indicates the changes of its length (top) or area (bottom). This shows the transient increase of the length and area of the middle and outermost contours during stage 1 as they are rotated and translated along the surface. The arrows show the direction of the movement of the ventralmost point of the contours, with the green and turquoise regions showing the positions of these points during stage 1 and stage 2, respectively.\vspace{-20pt}}
\label{sfig:SF_4}
\end{figure*}
\phantom{.}
\subsection{Supplemental Movie}
\noindent\textbf{Movie S1.} Contour dynamics. Contours are shown updating in time with smoothed nuclear positions visible as points. The nuclei are colored by the contour to which they belong. The movie shows the first 20 minutes of gastrulation.\label{movieS1}
\twocolumngrid
\bibliography{arxiv_ref}

\end{document}